\documentclass{emulateapj}     
\usepackage{color}

\begin{document}


\title{Gemini Follow-up of two massive {\sc H\,i} clouds discovered with the Australian Square Kilometer 
Array Pathfinder}


\author{Juan P. Madrid\altaffilmark{1}, Karen Lee-Waddell\altaffilmark{1},
Paolo Serra\altaffilmark{1,2}, B\"arbel S.\ Koribalski\altaffilmark{1},
Mischa Schirmer\altaffilmark{3},\\ Kristine Spekkens\altaffilmark{4},
Jing Wang\altaffilmark{1}}

\altaffiltext{1}{CSIRO Astronomy and Space Science, PO BOX 76, Epping NSW 1710, Australia}
\altaffiltext{2}{Osservatorio Astronomico di Cagliari, Instituto Nazionale di Astrofisica, Italy}
\altaffiltext{3}{Gemini Observatory, Southern Operations Center, Colina El Pino s/n, La Serena, Chile}
\altaffiltext{4}{Department of Physics, Royal Military College of Canada, PO Box 17000, Station Forces, Kingston ON K7K 7B4, Canada}

                         
\begin{abstract}

Using the Gemini Multi Object Spectrograph (GMOS) we search for optical counterparts 
of two massive  ($\sim10^9$M$_{\odot}$) neutral hydrogen clouds near the spiral galaxy
IC\,5270, located in the outskirts of the IC\,1459 group. 
These two {\sc H\,i} clouds were recently discovered using the Australian Square Kilometer 
Array Pathfinder (ASKAP).
Two low surface brightness optical counterparts to one of these  {\sc H\,i} clouds are identified 
in the new  Gemini data that reaches down to magnitudes of $\sim$27.5 mag in the $g$-band.
The observed {\sc H\,i} mass to light ratio derived with these new data, $M{_{\sc H\,I}}/L_g =242$,  
is among the highest reported to date. We are also able to rule out that the two {\sc H\,i} clouds 
are dwarf companions of IC 5270. Tidal interactions and ram pressure stripping are plausible 
explanations for the physical origin of these two clouds.

\end{abstract}


\maketitle

\section{Introduction}

In the standard hierarchical clustering model of galaxy formation large 
structures form from the accretion and merger of small dark matter halos 
\citep{white1978}. One of the main challenges of the $\Lambda$CDM model 
is the prediction of a large number of dark matter halos and satellite 
galaxies that remain observationally undetected. This limitation of the 
hierarchical clustering model could be solved by the detection of 
halos where star formation has been truncated so they remain optically faint
but detectable through neutral hydrogen ({\sc H\,i}) emission 
\citep[e.g.][]{bonamente2008}.  

Despite the impact that observations of pure  {\sc H\,i} proto-galaxies 
would have on cosmology, such detections remain rare. The main all-sky surveys of neutral 
hydrogen such as the {\sc H\,i} Parkes All Sky Survey \citep[HIPASS;][]{koribalski2004,meyer2004}
and the Arecibo Legacy Fast ALFA \citep[ALFALFA;][]{haynes2011} have found few examples of 
massive {\sc H\,i} detections without optical counterparts.


Among the 1000 {\sc H\,i}-brightest HIPASS sources \citep{koribalski2004}
HIPASS J0731--69 stands out as the only extragalactic source without a clear 
optical counterpart. HIPASS J0731--69, however, has a clear link with NGC 2442, a 
spiral galaxy with asymetric spiral arms and disturbed morphology located in 
the NGC 2434 group. The {\sc H\,i} reservoir that constitutes HIPASS J0731--69 
was most likely dislodged from NGC 2442 during a recent tidal encounter \citep{bekki2005}. 

In the northern hemisphere, of the 15,855 sources detected by 
the 40\% ALFALFA catalog only 1.5\% lack survey-grade optical counterparts 
\citep{haynes2011,cannon2015}. Out of the 15,855 sources of the 
40\% ALFALFA catalog 15,041 of them are extragalactic.
Recent observing campaigns targeting {\sc H\,i} sources without optical counterparts
in the ALFALFA survey have revealed ultra-low surface brightness structures, tidal 
dwarf galaxies and objects with exceptionally high mass-to-light ratio \citep{leewaddell2014,janowiecki2015,cannon2015}.

Other intergalactic {\sc H\,i} clouds without clear optical counterparts 
have also been discovered outside the main all sky surveys. For instance, 
\citet{osterloo2005} report a large {\sc H\,i} tail
($3.4\times10^8$M$_{\odot}$) near the center of the Virgo cluster. The origin of this 
cloud is likely ram pressure stripping of NGC 4388 by M\,86. Another example of 
intergalactic {\sc H\,i}, also in Virgo,  is {\sc H\,i} $1225 +01$ \citep{giovanelli1989,chengalur1995}.
{\sc H\,i} $1225 +01$ has two main components separated by 100 kpc with {\sc H\,i} 
masses of $10^9$M$_{\odot}$. One of the two components is associated with a low surface 
brightness galaxy while the second component does not have an identified optical 
counterpart \citep{matsuoka2012}. See also the example of the Vela cloud, a large and 
massive intergalactic {\sc H\,i} cloud residing in the NGC 3256 group \citep{english2010}.

The Leo Triplet (NGC 3623, NGC 3627, and NGC 3628) is the classical example of a 
large tail of neutral hydrogen \citep{haynes1979} where a faint tidal tail of 
stars, coincidental with the {\sc H\,i} peaks, was discovered with deep optical 
follow-up \citep{chromey1998}.

Similarly, in this paper, we use new deep Gemini images to search for the first signs of 
star formation within two massive {\sc H\,i} clouds recently discovered near the 
spiral galaxy IC\,5270. The two {\sc H\,i} clouds that are the target of this study have 
neutral hydrogen masses comparable to the Milky Way's.

\begin{deluxetable*}{cccccccc}
\tablecaption{Summary of observational properties\label{tbl-1}} 
\tablehead{
\colhead{Target Cloud} &  \colhead{R.A.} & \colhead{Dec} & \colhead{{\sc H\,i} mass} & \colhead{Observing} & \colhead{IQ} & \colhead{Cloud} & \colhead{Total exposure}\\
\colhead{} & \colhead{} & \colhead{} & \colhead{(M$_{\odot}$)} & \colhead{Date} & \colhead{} & \colhead{Cover} & \colhead{time (s)}\\
\colhead{(1)} & \colhead{(2)} & \colhead{(3)} & \colhead{(4)} & \colhead{(5)} & \colhead{(6)} & \colhead{(7)} & \colhead{(8)}\\
}

\startdata

North        &   22:57:54.64   & -35:47:59.1  & $1.6 \pm 0.4 \times10^9 $ & 2016 Aug 04   & 85\%     & 50\%  & 3000 \\
North East   &   22:57:54.64   & -35:47:59.1  & $1.0 \pm 0.2 \times10^9 $ & 2016 Aug 02   & 70-85\%  & 50\%  & 2200 \\
 \enddata

\tablecomments{Column (1): Target cloud; Column (2) and (3): R.A.\ and Dec of peak {\sc H\,i} intensity;
Column (4):  {\sc H\,i} mass in solar masses;  Column (5) Observing Date; Column (6): 
Image quality in percentile; Column (7): Cloud cover in percentile; Column (8): 
Total exposure time in seconds.}

\end{deluxetable*}


\section{ASKAP Observations of the IC 1459 group by Serra et al. (2015)}

The recent analysis of early commissioning data obtained with the Australian 
Square Kilometer Array Pathfinder (ASKAP) revealed the existence of three
galaxy-size clouds of neutral hydrogen in the galaxy group IC 1459
without any documented optical emission \citep{serra2015}. 
One of these clouds is linked to NGC 7418 while the two other clouds, which are the 
target of this study, are located near IC\ 5270. Each of these two large {\sc H\,i} 
accumulations have the characteristic  {\sc H\,i} mass of galaxies ($\sim10^9$M$_{\odot}$) 
but, intriguingly, have no optical counterpart in the Digital Sky Survey -- 
see Figure 1, top panel.

The galaxy group IC 1459 is located at a distance of 29 Mpc \citep{blakeslee2001}
and is named after its brightest ($m_B =11.0$ mag), early type member. 

The 11 ASKAP detections with optical counterparts in this galaxy group span 
a range of {\sc H\,i} masses from $5.4\times 10^9$ M$_{\odot}$ down to 
$0.5\times 10^9$ M$_{\odot}$. The two {\sc H\,i} clouds we follow up with the 
Gemini  telescope stand out as they are more massive in {\sc H\,i} than 
several of the other detections in IC 1459 which have bright optical 
counterparts.

The {\sc H\,i} emission associated with IC 5270 has a heliocentric  velocity of 
$v_{hel}\sim 1790-2110$ km/s in the  ASKAP spectra obtained by \citet{serra2015}, 
in agreement with the optical redshift of 1858 km/s \citep{dacosta1998}. 
The two  {\sc H\,i}  clouds have emission at radial velocities of $v_{hel}\sim 1900-2110$ km/s. 
The HIPASS data of this system shows that IC 5270 and the two neighbouring 
hydrogen clouds are linked by a diffuse and extended {\sc H\,i} envelope.

The largest of the two clouds (North cloud) has a slight velocity gradient 
in {\sc H\,i} -- \citet{serra2015}, their figure 6. The second, lighter, North-East 
cloud  has a far more homogeneous velocity field (moment-one) in the ASKAP data. 
The HIPASS name for the IC 5270 system (galaxy + clouds) is
HIPASS J2258--35.


\section{Gemini Observations and Data Reduction}

We obtained deep $g$-band images of these two {\sc H\,i} clouds with the Gemini
Multi Object Spectrograph (GMOS) on Gemini South under program GS-2016B-Q-61. 
Imaging in the $g$-band (475 nm) was chosen given that this filter is sensitive to both 
old and new stellar populations \citep[e.g.][]{leewaddell2014}. GMOS was used with 
2x2 binning, resulting in a plate scale of $0.16\arcsec$ pixel$^{-1}$.

\begin{figure*}
        \centering
        \label{tab:example_table}
        \begin{tabular}{cc} 
                \includegraphics[scale=0.48]{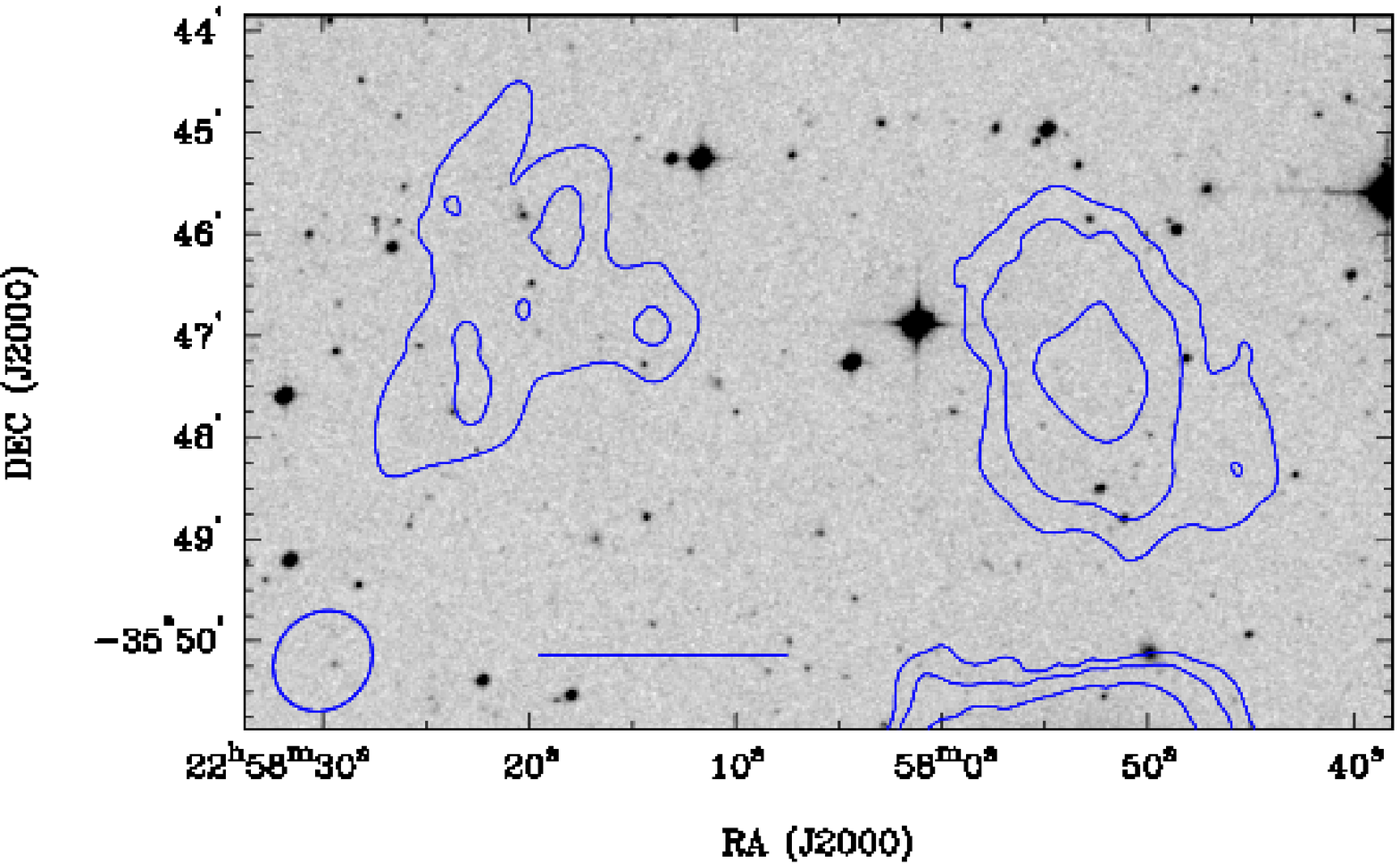}\\                
                \includegraphics[scale=0.48]{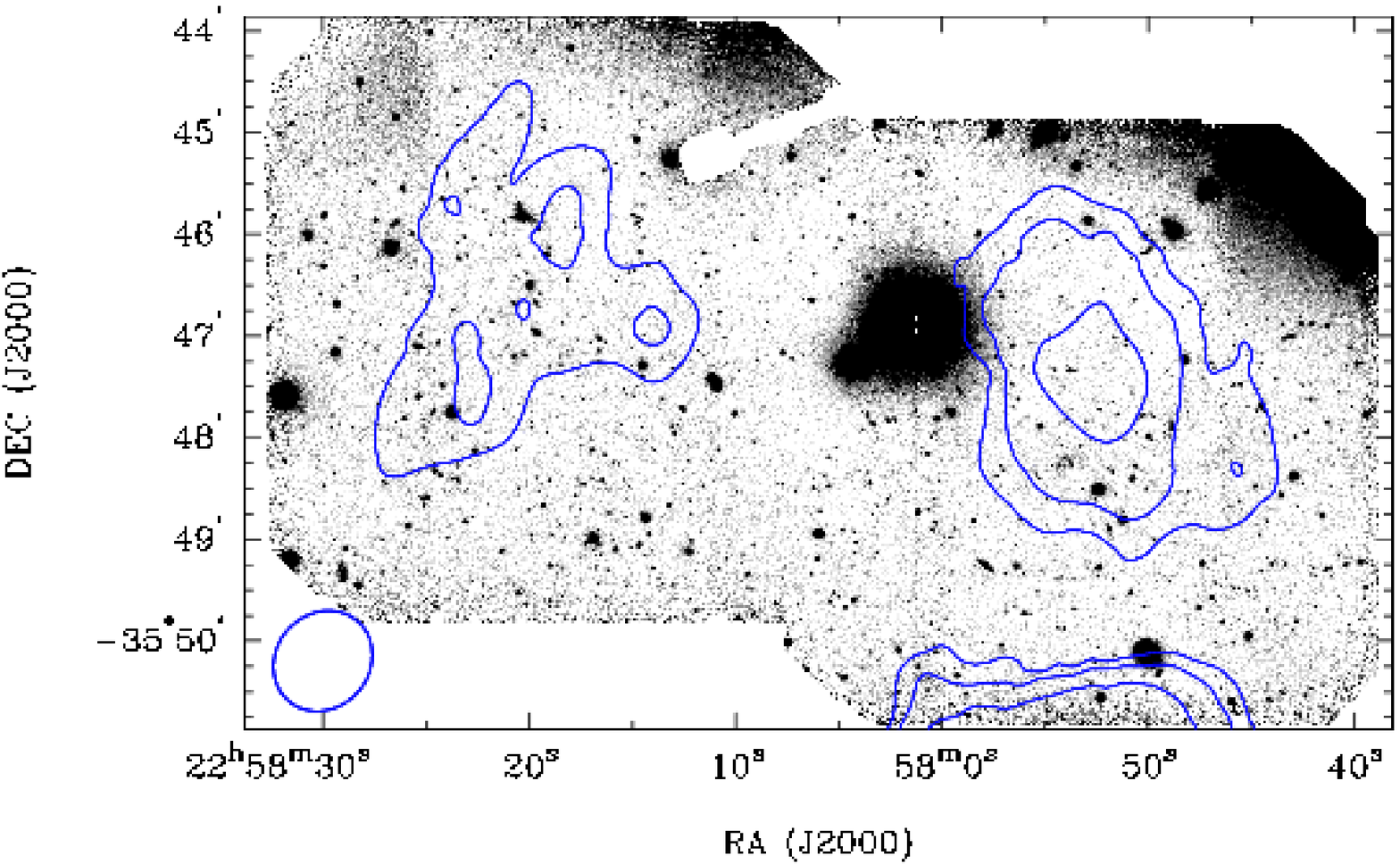}\\
                \includegraphics[scale=0.48]{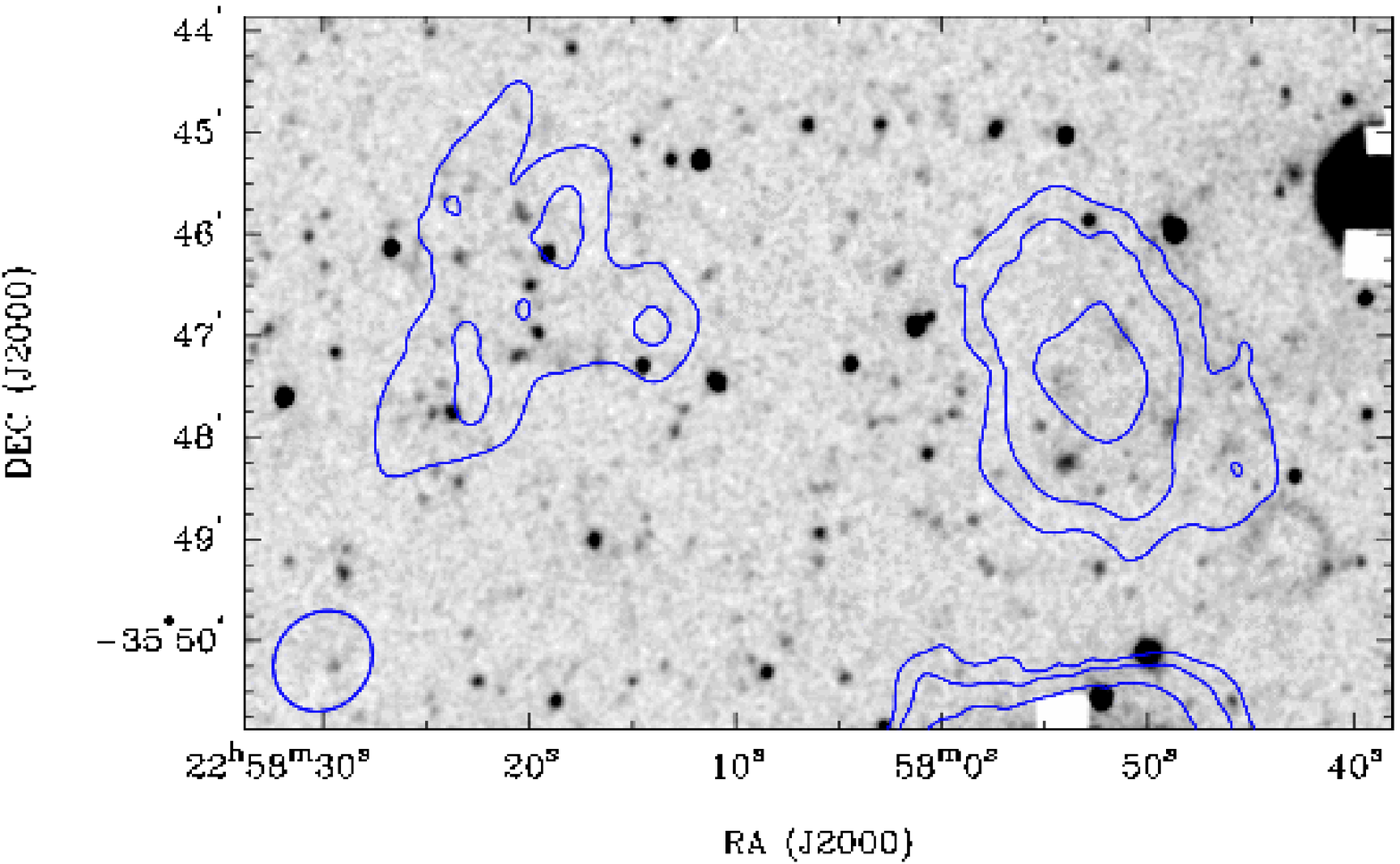}\\
        \end{tabular}
    \caption{{\it Top panel:} Blue ASKAP contours of the two {\sc H\,i} clouds
located North and North-East of IC 5270 overlaid on a Digital Sky Survey (red) 
image. The contours visible on the lower right hand corner are those of IC 5270.
The physical size of this image is $\sim50\times85$ kpc, the scale bar represents $\sim$20 kpc.
North is up and east is left. The radial velocity range of the {\sc H\,i} emission is $v_{hel}\sim 1900-2110$ km/s. 
{\it Middle panel:} Mosaic of the Gemini GMOS $g$-band images of the same area 
of the sky as the above panel. The vignetting is due to the placement of the 
On Instrument Wavefront Sensor (OIWFS) on a guide star within the science field. 
{\it Bottom panel:} Archival Near-Ultraviolet GALEX image. This GALEX image was smoothed 
by a Gaussian kernel of 2 pixels.\\}
\end{figure*}

\begin{figure*}
        \centering
        \label{tab:example_table}
        \begin{tabular}{cc}
                \multicolumn{2}{c}{\includegraphics[scale=0.6]{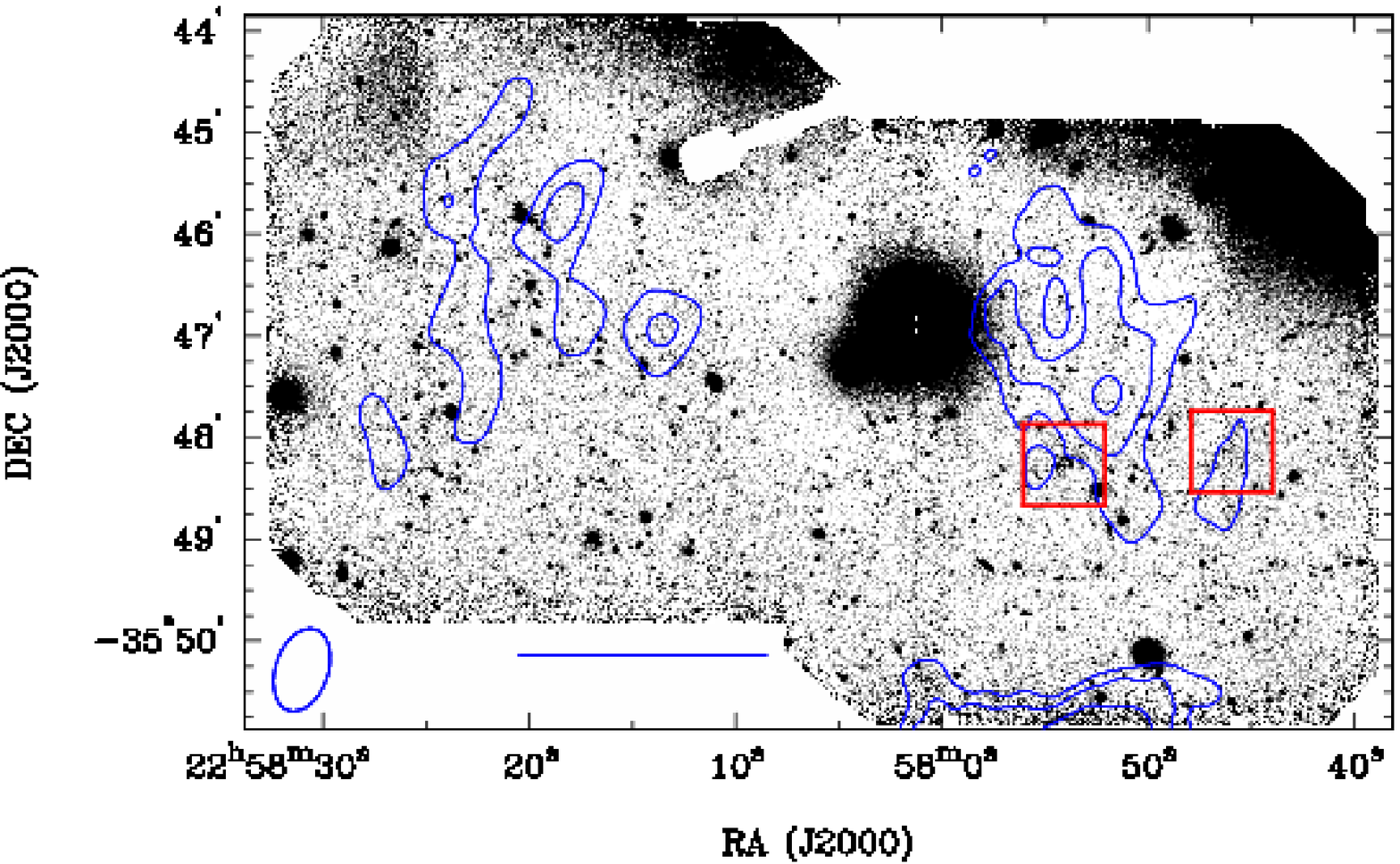}}\\                
                \includegraphics[scale=0.3]{geminiv35.ps} & \includegraphics[scale=0.3]{geminiv45.ps}\\
                \includegraphics[scale=0.3]{galexv35.ps} & \includegraphics[scale=0.3]{galexv45.ps}\\           
        \end{tabular}
    \caption{{\it Top panel:} New Gemini image with ASKAP contours obtained with the new data reduction 
presented here which trace the regions with the highest {\sc H\,i} column densities. The highest column density on
the North cloud is $N_{H}=1.4\times10^{21}$cm$^{-2}$ while its outer contours trace column densities of 
$N_{H}\sim 0.5\times10^{21}$cm$^{-2}$. The highest column density of the North-East cloud is $N_{H}=0.7\times10^{21}$cm$^{-2}$.
The red boxes show the location of the two optical counterparts. The scale bar represents $\sim$20 kpc.
{\it Middle panel left:} Zoom of the Gemini image of the western optical counterpart (ID 1 on Table 2).
{\it Middle panel right:} Zoom of the Gemini image of the western optical counterpart (ID 2 on Table 2). 
{\it Bottom panel left:} Zoom of the GALEX image of the eastern optical counterpart (ID 1 on Table 2).
{\it Bottom panel right:} Zoom of the GALEX image of the eastern optical counterpart (ID 2 on Table 2).
The archival GALEX image has an exposure time of 3306 seconds and was observed in 2010 October 31.\\  
}
\end{figure*}



The two targets were observed on 2016 August 02 and 2016 August 04 under photometric conditions 
(Cloud Cover 50th percentile). Importantly for our science objectives, all observations 
were obtained with the darkest sky conditions (Surface Brightness 20th percentile). 
These observing conditions are optimal to detect extended low-surface brightness objects 
in $g$-band. The seeing ranged between 0.9$\arcsec$ and 1.5$\arcsec$, resulting in an effective 
image seeing in the coadded image of 1.1$\arcsec$. Fourteen exposures of 200 s were obtained for 
the North cloud, and eleven exposures of the North-Eastern cloud. 

A 20$\arcsec$ wide dither pattern was chosen to cover the gaps between the three GMOS detectors, 
and the two GMOS pointings overlap so that we could image a continuous field over the area of 
the {\sc H\,i} detections. A summary of the observations is given in Table 1. 

Data were reduced with the THELI pipeline \citep{schirmer2013,erben2005} following 
standard procedures. Exposures were overscan corrected, bias subtracted, and flat fielded 
using twilight exposures. Individual weight maps were created to mask cosmic rays and 
other detector defects. For the astrometric calibration, THELI uses {\tt Scamp} \citep{bertin1996}. 
Source catalogs (mostly stars) were extracted from the images and matched against the Initial 
Gaia Source List \citep[IGSL,][]{smart2014} in the J2000 ICRF reference frame. Images were 
distortion-corrected using a third-order polynomial description, and registered with respect 
to each other with a typical precision of 1/20th pixel.

Accurate, yet conservative background correction of individual exposures is crucial for the 
detection of extended low-surface brightness objects. Background subtraction was carried out 
several times and with different parameters in order to ensure that faint structures both small and 
large were detectable. THELI uses {\tt SExtractor} \citep{bertin1996} to detect 
and mask objects in the field; {\tt SExtractor} was configured with a detection threshold of 1.5 $\sigma$ 
and a minimum number of 10 connected pixels above the detection threshold. The size of the object masks was 
extended by a factor of three (see Schirmer et al.\ 2015) to include hidden flux around faint objects. 
The images that we present in Figures 1 and 2, were created using the masks described above that were
then convolved with a ${\rm FWHM}=400$ pixel wide ($1.1^{\prime}$ Gaussian kernel, and the resulting 
background models were subtracted. For the data reduction presented in Figures 1 and 2 all 
potential structures with an angular extent of less 
than $\sim1^{\prime}$ (8 kpc physical scale) are  preserved. Several trials of background subtraction 
with different parameters were carried out to ensure that very faint structures of all sizes and with 
fluxes similar to the background level were preserved. 


At the distance of IC 5270 (29 Mpc) one arcsecond corresponds to a 
physical scale of 137 pc \citep{wright2006}. The GMOS field of view of 5.5$\times$5.5 
arcmin$^2$ thus corresponds  to an area of $\sim$45.2$\times$45.2 kpc$^2$.
On our coadded image we can see structures of 150 pc at the distance of IC 5270, 
significantly smaller than typical Tidal Dwarf Galaxies \citep{leewaddell2016}.
A mosaic of the two fields imaged with GMOS is presented in Figure~1 -- middle panel.

\section{Results}

The most evident result, clearly visible in Figure 1, is the lack of both 
large diffuse optical counterparts and of stellar streams in the field of the
two {\sc H\,i} clouds. The new Gemini data, while deep enough to reveal the presence 
of stellar tidal tails, like the ones found by \citet{leewaddell2012}, is free of 
such large scale features. With the Gemini images we detect the faint outer isophotes 
of the IC\ 5270 disk but no tidal disturbance is found on the northern part of this galaxy.  
We are also able to rule out that the two {\sc H\,i} clouds are dwarf companions of IC 5270.

On the Gemini images we identify two ultra-faint optical counterparts to the North Cloud.
The overall morphology of these objects is similar to the sources recently presented by 
\citet{leisman2017}. That is, clusters of clumps associated with ultra-diffuse emission.
Moreover, these sources also have  detectable flux in Near 
Ultraviolet (NUV) archival images obtained by GALEX. The NUV is a very sensitive tracer 
of recent star formation. The location of these ultra-low surface brightness structures
is highlighted on the top panel of Figure 2 with two red boxes.
A zoom on these sources is shown in Figure 2 where the new Gemini images (middle panels) 
are presented alongside the archival GALEX images (bottom panels).

The top panel of Fig.\ 2 shows the new Gemini image and the {\sc H\,i} contours 
of a new data reduction of the ASKAP data. The new data reduction of the radio data 
was carried out giving more weight to longer baselines and improving our resolution,
that is, using robust~=0. For the robust = 0 weighting the PSF is $48\arcsec \times 35\arcsec$ 
and the peak {\sc H\,i} brightness 
in the North cloud is $\sim2.2$ Jy beam$^{-1}$ km s$^{-1}$  which translates into $N_{H}=1.4\times10^{21}$cm$^{-2}$.
This new data reduction was carried out in order to identify those regions with the highest 
{\sc H\,i} column density, where star formation has the highest likelihood to occur.

With our new Gemini data we are able to detect sources down to $g$-band magnitudes 
of $\sim$27.5 mag, at three sigma above the local background. By running {\tt SExtractor} 
with a 3$\sigma$ detection threshold on the deep Gemini images we detect $\sim$ 1900 sources within 
the field of view. This is in contrast with the much shallower DSS image that 
appears virtually free of sources in the area of interest. Indeed, when we run 
{\tt SExtractor} on the DSS image (top panel Fig.\ 1) only 128 sources are detected.

Two bright objects are easily identifiable within the {\sc H\,i} contours of IC\ 5270.
On the DSS and GALEX images these two objects are the only ones present in the 
northern edge of IC\ 5270. The brightest object (RA=22:57:49.94, Dec=-35:50:07.9) reveals 
itself as a background face-on spiral on the Gemini data. The second object 
(RA=22:57:52.29, Dec=-35:50:33.4) is ultra-compact and it is likely to be a background 
star-forming galaxy.


\subsection{Photometry of faint optical counterparts}

We carry out aperture photometry of the faint optical counterparts with the task \texttt{phot} 
within the \texttt{daophot} package in \texttt{Pyraf}. We measure the flux of the optical 
counterparts with an aperture of 35 and 70 pixels in radius, as a function of their 
respective sizes. The eastern source (ID 1 on Table 1) is indeed more compact than the source 
to the west  (ID 2 on Table 1). We use a zeropoint 
of $ZP = 27.32$ mag to convert fluxes to  $g$ band magnitudes. We obtain m$_g$ =21.3 $\pm 0.25$ mag 
 for the source on the left of Figure 2, middle panel (ID 1) and m$_g$ = 21.2 $\pm 0.25$ mag for 
the other source (ID 2).
By assuming that the neutral hydrogen clouds and the optical counterpart are at the same 
distance as IC\ 5270 (29 Mpc) we derive absolute magnitudes of $M_g =-11.01$ mag 
and $M_g =-11.15$.

By adopting the absolute magnitude of the Sun to be M$_{\odot,g}$ =5.12 mag 
\citep{sparke2007} we can obtain the solar luminosity of the ultra faint optical 
counterparts of the North cloud to be $L=3.1 \pm 0.6 \times 10^6 L_{\odot}$ and 
$L=3.5 \pm 0.66 \times 10^6 L_{\odot}$. Properties of the ultra-faint and ultra-diffuse 
optical counterparts are summarized in Table 2.

\subsection{An extremely high {\sc H\,i} mass to stellar light ratio}

With the absolute magnitudes obtained with the new Gemini data and the {\sc H\,i}
masses published by \citet{serra2015} we can now derive an observed {\sc H\,i} 
mass-to-light ratio:

\begin{equation}
\label{eqn:abs_mag}
\frac{M_{\sc H \sc{I}}}{L_g} =\frac{1.6 \times 10^9}{6.6 \times 10^6}=242 \pm 76.
\end{equation}	 

This gas-to-light ratio is higher than other extreme values recently
published:  \citet{janowiecki2015} give a value of  $M_{\sc H \sc{I}}/L_g > 57$ 
for the HI1232+20 system and  \citet{janesh2017} find an upper limit of 144 for the gas
to light ratio of AGC 249525. 

The North-East cloud lacks any obvious optical counterpart on our deep Gemini images.
We derive a crude lower limit of the gas-to-light ratio for the North-East cloud
by assuming that any existing optical counterpart will be fainter than the faintest detection
we present here:

\begin{equation}
\label{eqn:abs_mag}
\frac{M_{\sc H \sc{I}}}{L_g} =\frac{1.0 \times 10^9}{3.1 \times 10^6} > 322 \pm 101.
\end{equation}	


\begin{deluxetable}{ccccc}
\tablecaption{Properties of Optical Counterparts \label{tbl-1}} 
\tablehead{
\colhead{ID} & \colhead{R.A.} & \colhead{Dec} & Absolute $g$& \colhead{$L_{\odot}$}\\
\colhead{} & \colhead{} & \colhead{} & \colhead{Magnitude} & \colhead{}\\
\colhead{(1)} & \colhead{(2)} & \colhead{(3)} & \colhead{(4)} & \colhead{(5)}
}
\startdata

(1)      & 22:57:54.00   & -35:48:15.6  & -11.01 & 3.1$\times 10^6$\\
(2)      & 22:57:45.92   & -35:48:04.7  & -11.15 & 3.5$\times 10^6$\\

 \enddata

\tablecomments{ Column (1): Optical counterpart ID, counterpart with ID 1 is the easternmost 
object, counterpart with ID 2 is the westernmost one; 
Column (2): Right ascension;
Column (3): Declination;
Column (4): Absolute magnitude in $g$-band;
Column (5): Equivalent solar luminosities.\\}

\end{deluxetable}


\section{Discussion}

\subsection{``Two missing galaxies problem"}

As mentioned in the Introduction, the two neutral hydrogen clouds that we follow up with Gemini
have similar {\sc H\,i} content as the Milky Way. One possibility to explain the absence of 
two Milky-Way-sized galaxies cospatial with these clouds is their low {\sc H\,i} column density. 
The surface density threshold for star formation is given by \citet{schaye2004} in terms of {\sc H\,i} 
column density: $N_{H,crit}\sim (3-10)\times10^{20}$cm$^{-2}$. This value is in agreement with the 
column density required to detect clustered star formation as given by \citet{mullan2013}.

The new data reduction carried out for this work shows that the maximum
{\sc H\,i} column density is above the expected critical 
threshold to trigger star formation. Given that relatively few stars 
are detected compared to the amount of available {\sc H\,i} makes this
system an extreme case of star formation inefficiency. However, the derived column density might be above 
the star formation threshold only in projection and the {\sc H\,i} might be too diffuse to have significant 
star formation. Assuming a constant uniform distribution of {\sc H\,i} in a 8$\times$8$\times$8 kpc cube, 
the volume density of the brightest {\sc H\,i} features would be 0.06 cm$^{-3}$, within the range 
of the star formation threshold found by \citet{schaye2008}: 0.01 cm$^{-3}$ to 0.1 cm$^{-3}$ but 
below higher values also commonly used \citep[0.13 cm$^{-3}$, e.g.][]{dave2017}. Instead of
a uniform distribution, the {\sc H\,i} is more likely to form sheets, filaments and dense clumps 
where pockets of star formation can occur.



\subsection{Origin of the {\sc H\,i} clouds}

From the analysis of ASKAP and HIPASS data \citet{serra2015} concluded that the
two massive {\sc H\,i} clouds must be `the tip of the iceberg' of a large scale
distribution of {\sc H\,i} gas that encompasses IC 5270. 

Is the origin of these clouds the result of tidal interactions between IC 5270 and
a nearby group member? As mentioned above, we detect the faint 
outer isophotes of the IC\ 5270 disk on the new Gemini images but no streams 
or tails are detected. An inspection of IC 5270 stellar morphology of IC 5270 reveals 
that this galaxy has smooth isophotes. The surface brightness profile 
of IC 5270 in the DSS2 red filter is smooth and it is well fit by a single S\'ersic model. 
The DSS2 blue filter shows a central component on the surface brightness profile that 
is likely due to star formation. In any event, the optical data available for IC 5270 
show no signs of strong tidal interactions. However, mild tidal interactions between group 
members can create starless {\sc H\,i} tails by removing gas-rich material from the
outskirts of disk galaxies \citep{bekki2005} without affecting the distribution of
stars. The large number of {\sc H\,i} `debris' present within the IC 1459 galaxy group
are signposts of rich interactions between group members \citep{saponara2018}. 

Ram pressure stripping by the intergroup medium offers another plausible explanation for 
the origin of these two clouds. Indeed, ram pressure stripping can remove gas from 
galaxies that are either not tidally interacting or experiencing weak tidal 
interactions \citep{hester2006}. Ram pressure could have dislodged 
the {\sc H\,i} material that constitutes these two clouds from the outskirts of IC\ 5270
without altering its stellar structure. Limitations to the ram pressure hypothesis are
the facts that the IC 1459 is a group with low velocity dispersion $\sigma =223 \pm 62$ km/s 
\citep{osmond2004} and IC 5270 is more than 400 kpc from the group center.

\acknowledgments


Based  on observations  obtained at  the Gemini  Observatory  which is 
operated by the AURA under a cooperative  agreement with  the NSF  on 
behalf  of the Gemini partnership.


The Australian SKA Pathfinder is part of the Australia Telescope National Facility 
which is managed by CSIRO. Operation of ASKAP is funded by the Australian 
Government with support from the National Collaborative Research Infrastructure 
Strategy. ASKAP uses the resources of the Pawsey Supercomputing Centre. 
Establishment of ASKAP, the Murchison Radio-astronomy Observatory and the Pawsey 
Supercomputing Centre are initiatives of the Australian Government, with support from the 
Government of Western Australia and the Science and Industry Endowment Fund. 
We acknowledge the Wajarri Yamatji people as the traditional owners of the 
Observatory site.

This project has received funding from the European Research Council (ERC) 
under the European Union's Horizon 2020 research and innovation programme (grant 
agreement no. 679627: FORNAX).



\end{document}